\documentclass[a4paper,amsfonts,amssymb,twocolumn,showpacs,showkeys,nofootinbib,floats,pre,final]{revtex4}
\usepackage{amsfonts}
\usepackage{amsmath}
\usepackage{amssymb}
\usepackage{graphicx}

\begin{document}

\title{Weak chaos and metastability in a symplectic system of many long-range-coupled standard maps}

\author{Luis G. Moyano}
\email[e-mail address: ]{moyano@cbpf.br}
\affiliation{Centro Brasileiro de Pesquisas F\'{\i}sicas \\
Rua Xavier Sigaud 150, \\
22290-180 Rio de Janeiro--RJ, Brazil }
\author{Ana P. Majtey}
\email[e-mail address: ]{amajtey@famaf.unc.edu.ar}
\affiliation{Facultad de Matem\'{a}tica, Astronom\'{\i}a y
F\'{\i}sica, Universidad Nacional de C\'{o}rdoba   \\
Ciudad Universitaria, 5000, C\'{o}rdoba, Argentina\\ CONICET}
\author{Constantino Tsallis}
\email[e-mail address: ]{tsallis@santafe.edu} \affiliation{Santa Fe Institute \\
1399 Hyde Park Road, Santa Fe, NM 87501, USA\\
and \\
Centro Brasileiro de Pesquisas F\'{\i}sicas \\
Rua Xavier Sigaud  150, 22290-180 Rio de Janeiro--RJ, Brazil
}
\date{\today}

\begin{abstract}
We introduce, and numerically study, a system of $N$ symplectically and globally coupled
standard maps localized in a $d=1$ lattice array. The global coupling is modulated
through a factor $r^{-\alpha}$, being
$r$ the distance between maps. Thus, interactions are {\it long-range} (nonintegrable) when
$0\leq\alpha\leq1$, and {\it short-range} (integrable) when $\alpha>1$.
We verify that the largest Lyapunov exponent $\lambda_M$ scales as $\lambda_{M} \propto
N^{-\kappa(\alpha)}$, where $\kappa(\alpha)$ is positive when interactions are
long-range, yielding {\it weak chaos} in the thermodynamic
limit $N\to\infty$ (hence $\lambda_M\rightarrow 0$). In the short-range case,
$\kappa(\alpha)$ appears to vanish,
and the behaviour corresponds to {\it strong chaos}. We show that, for certain
values of the control parameters of the system, long-lasting metastable states
can be present. Their duration $t_c$ scales as $t_c \propto N^{\beta(\alpha)}$,
where $\beta(\alpha)$ appears to be numerically consistent with the following
behavior: $\beta >0$ for $0 \le \alpha < 1$, and zero for $\alpha\ge 1$.
All these results exhibit major conjectures formulated within nonextensive statistical
mechanics (NSM). Moreover, they exhibit strong similarity between the present
discrete-time system, and the $\alpha$-XY Hamiltonian ferromagnetic model,
also studied in the frame of NSM.

\end{abstract}


\maketitle

\section{Introduction}
\label{intro}

The study of complex systems has been lately one the most active
areas of investigation. This multidisciplinary subject has
interest in natural sciences as well as in social and artificial
systems. In physics, a major area involved is
statistical mechanics. It is precisely for such systems that
nonextensive statistical mechanics (NSM)~\cite{tsallis_01} appears
to have its applicability. Typical features present in complex systems are
long-range interactions, long-term memory, fractal phase-space
structure,  scale-free network structure, or even combinations of these
characteristics. Quite frequently these systems cannot be correctly described
by the well-established Boltzmann-Gibbs
statistical mechanics (BGSM). Indeed, they often fail to verify its
basic assumptions (equiprobability of phase-space
occupation~\cite{boltzmann} and ergodicity~\cite{kinchin} at the thermal equilibrium state). NSM
generalises the usual BGSM formalism through the definition of the
q-entropy $S_q \equiv \frac{1-\int dx \, [p(x)]^q}{q-1}$ (with $S_1 =
S_{BG} \equiv -\int dx \, p(x) \ln p(x)$, where $BG$ stands for {\it
Boltzmann-Gibbs}) under simple constraints~\cite{prato}.

This theory has been extensively applied to nonlinear dynamical
systems~\cite{tsallis_01,tsallis_02,latorabaranger_02,applications}, as they are one of the most
widespread and useful ways for modelling complex phenomena.
For example, Hamiltonian systems are at the core of physics and consequently their relevance is
evident. There has been an important number of results indicating that certain
conservative models (e.g., the $\alpha$-XY
model~\cite{antoniruffo,anteneodotsallis}, the $\alpha$-Heisenberg
model~\cite{nobretsallis1,nobretsallis2}, the $\alpha$-Lennard-Jones-like
gas~\cite{borgestsallis}) can present a behaviour departing from the one predicted
by the BG formalism. For certain classes of initial conditions and parameters,
the system is prevented from virtually attaining its expected equilibrium state in
finite time when $N \to \infty$. In other words, if
the $N\to\infty$ limit is taken before the $t\to\infty$ limit, the system
becomes nonergodic. This nonergodicity is also reflected in a
variety of anomalous behaviours such as non-Gaussian momenta probability density
functions, negative specific heat, aging, and others~\cite{latora_01,cannastamarit,tamarit,cabral}.
On the other hand, other (simpler) nonlinear dynamical systems, as for example
maps, have emerged in many contexts, very often exhibiting new and interesting
results. Among these systems we find the logistic map, the standard map, coupled map lattices, discretised
Lorentz gas, and many others~\cite{ott}.
Both low- and high-dimensional discrete-time dynamical systems have been studied under the
framework of NSM, exhibiting various connections with Hamiltonian
dynamics~\cite{tsallis_03,moyanoetal,baldovinetal}.

It is known quite well that the number of degrees of freedom of a system defines the
possibilities that its dynamics may approach. A clear illustration of this fact in low-dimensional
systems (e.g., the 2-dimensional Chirikov-Taylor standard map) is the KAM tori, a
complex structure in the map phase-space that separates chaotic from
regular regions.
While much is known for systems with few degrees of freedom, the situation
is more intricate when many elements are involved. This is because the
$\Gamma$ phase-space grows in an extremely rapid manner.
In addition to this, studying continuous-time many-body systems can turn out to be very
difficult, if not impossible at all, due to the considerable
computational time needed to integrate the evolution equations.  This is even more
so when long-range forces are involved, as it is not justified to neglect
any interaction between the elements.

An alternative for this problem is the substitution of the continuous-time system for simpler
discrete-time systems, such as maps, which conserve many of the important
features of the physics involved.
This substitution is in fact justified interpreting a dimensional map
system as the intersection (in the Gibbs
$\Gamma$ phase-space) of a Poincar\'e plane with the orbit of a higher
dimensional Hamiltonian. More precisely, if we consider a
{\it time-independent} Hamiltonian system with ($N+1$) degrees of freedom
(($2N+2$)-dimensional $\Gamma$ phase-space), a ($2N$)-dimensional symplectic map is the
result of taking a Poincar\'e section over the constant energy hypersurface~\cite{ott}.
The recurrence time is discrete and the map is useful in displaying various
properties of the original Hamiltonian system.

NSM concepts have been shown to have an important role in low-dimensional
maps at the edge of chaos.
The archetypical logistic map (a $D=1$, dissipative,  nonlinear map) is one of the most
widely studied dissipative systems. In part due to its simplicity, it is
often used to illustrate many of the most important features of chaos.
In recent works, Robledo and Baldovin~\cite{baldovinrobledo} have analytically proved,
using standard renormalisation-group techniques, that the dynamics of the logistic map
at its critical point is unmistakenly well-described within a NSM frame. The sensitivity to
initial conditions is a $q$-exponential function~\cite{qexp}, and is
related to the entropy production through the $q$-generalised Pesin-like identity,
linking the sensitivity to initial conditions to the $q$-entropy $S_q$ with
$q=0.2445\ldots$. Moreover, the logistic-map with noise (a Langevin-like
generalisation of the usual logistic map) has been found to
present two-step relaxation processes and aging (presenting interesting common points
with slow glassy dynamics~\cite{robledoglassy}).
Other aspects of the NSM formalism were also studied in the more general
case of one-dimensional $z$-logistic family of maps considering the attractor
fractal nature~\cite{zlogistic}.
Regarding two-dimensional ($D=2$) {\it dissipative} systems (as the Hen\'on map or its
linearised version, the Lozi map), results
indicate that it presents the same value of $q$ as the
logistic map, therefore suggesting a common universality
class~\cite{tirnaklietal}. Regarding two-dimensional ($D=2$)
{\it conservative} maps, a very interesting example, the triangle Casati-Prosen
map (mixing, ergodic, but with vanishing Lyapunov exponent), has been recently
been studied in connection to the entropy $S_q$ with $q=0$
\cite{CasatiTsallisBaldovin}.

Moving further into conservative discrete-time systems, two-step relaxation has been
also observed at the edge of chaos for the Chirikov-Taylor standard map, a
paradigmatic one for $2$-dimensional
symplectic maps. This map has been thoroughly studied
and explained by means of the KAM-theorem~\cite{ott}. This map is known to be
area preserving, hence its variables are often considered as
``coordinate'' and ``momentum'', in analogy with Hamiltonian systems.
Along this lines, it is appealing to look to the more general case of a {\it symplectic} system of $N$
coupled standard maps ($D=2N, N \ge 1$). Different efforts were done for values of $N$ as low as $N=2$ up
to $N=500$~\cite{baldovinbrigatti,baldovinetal,latorabaranger_01}. In these cases, the
same two-step relaxation was found as well as other features pointing
to a NSM applicability. But in all these cases, the
coupling was done trough the {\it momentum} variables. Even though this has its
own interest, it would be instructive to see the effects of coupling such a
system through the {\it coordinate} variables, as in more realistic
situations.

In this work we will study a high-dimensional globally coupled conservative map system that,
as discussed above, presents many of the characteristics of Hamiltonian dynamics.
Our purpose is to contribute to the understanding of the role that NSM plays in
the anomalous features present in long-range-interacting dynamical systems.
Our results show interesting similarities between this map model and
many-body long-range-interacting Hamiltonian dynamics, in particular the $\alpha$-XY model. In both
cases, long-lasting metastable states, as well as weak chaos, are present in the
thermodynamic limit.
Both features point out that under certain initial conditions, orbits do {\it not}
visit with equal probability the entire $\Gamma$ phase space, or in other words there is a failure
in ergodicity.

In the next section we introduce and describe the model, as well as some
relevant details of our simulations. In section \ref{results}
we present our numerical results. We report on the scaling behaviour of the
largest Lyapunov exponent with the system size $N$, and point out connections
with Hamiltonian dynamics. We analyse the relaxation to
equilibrium, and we make a systematic characterisation as a function of the system
parameters. Finally, a discussion and summary are presented in
section {\ref{conclusions}}.

\section{Model}
\label{model}
Our model is a set of $N$ symplectically coupled (hence conservative) standard
maps, where the coupling is made through the {\em coordinates} as follows:

\begin{eqnarray}
\begin{array}{llll}
\theta_i(t+1)& = & \theta_i(t) + p_i(t+1)& \;({\rm mod}\;1), \\
             &   &\mbox{}\\
p_i(t+1)     & = & p_i(t) + \frac{a}{2\pi}\sin[2\pi\theta_i(t)] +\\
             &   &\mbox{}\\
             &   &\frac{b}{2\pi\tilde N}\sum\limits_{\substack{
\begin{subarray}{1}
j=1\\j\neq i
\end{subarray}
}
}^N\frac{sin[2\pi(\theta_i(t)-\theta_j(t))]}{r^\alpha_{ij}}& \;({\rm mod}\;1),

\end{array}
\label{eq_azp}
\end{eqnarray}
where $t$ is the discrete time $t=1, 2, \ldots$, and $\alpha \ge 0$. The $a$ parameter
is the usual nonlinear constant of the individual standard map, whereas
the $b$ parameter modulates the overall strength of the
long-range coupling. Both parameters contribute to the
nonlinearity of the system; it becomes integrable when $a=b=0$. For simplicity,
we have studied only the cases $a>0$, $b>0$, but we expect similar results
when one or both of these parameters are negative. The systematic study of the
whole parameter space is certainly welcome.
Notice that, in order to describe a system whose phase space is bounded, we
are considering, as usual, only the torus (${\rm
mod}\;1$). Additionally, the maps are localised in a one dimensional ($d=1$) regular
lattice with periodic boundary conditions. The distance $r_{ij}$ is
the minimum distance between maps $i$ and $j$, hence it can take values from
unity to $\frac{N}{2}$ ($\frac{N-1}{2}$) for even (odd) number $N$ of
maps. Note that $r_{ij}$ is a fixed quantity that,
modulated with the power $\alpha$, enters Eq.~(\ref{eq_azp}) as an
effective time-independent coupling constant. As a consequence,
$\alpha$ regulates the range of the interaction between maps. The
sum is global (i.e., it includes every pair of maps), so the limiting
cases $\alpha=0$ and $\alpha=\infty$ correspond respectively to infinitely long range
and nearest neighbours. In our case $d$=1, thus
$0 \le \alpha \le 1$ ($\alpha>1$) means {\it long-range} ({\it short-range}) coupling. Moreover, the
coupling term is normalised by the sum ~\cite{kacetal,anteneodotsallis} \mbox{$\tilde
N \equiv d\int_1^{N^{1/d}}dr\; r^{d-1}\:r^{-\alpha}=
\frac{N^{1-\alpha/d}-\alpha/d}{1-\alpha/d}$}, to yield a
non-diverging quantity as the system size grows (for simplicity, we have
replaced here the exact discrete sum over integer $r$ by its continuous
approximation).

If $G(\bar x)$ denotes a map system, then $G$ is symplectic when its Jacobian
$\partial G/\partial \bar x$ satisfies the relation~\cite{ott}:
\begin{equation}
\left(\frac{\partial G}{\partial \bar x}\right)^TJ\left(\frac{\partial
G}{\partial \bar x}\right)=J \label{e.simplectic} \,,
\end{equation}
where the superindex $T$ indicates the transposed matrix, and $J$ is the Poisson matrix, defined by
\begin{equation}
J \equiv\left(\begin{array}{cc} 0 & I\\ -I & 0 \end{array} \right) \,,
\end{equation}
$I$ being the $N\times N$ identity matrix.
A consequence of Eq.~(\ref{e.simplectic}) is that the Jacobian determinant
$|\partial G /\partial \bar x|=1,$ indicating that the map $G$ is {\em (hyper)volume-preserving}.
In particular, for our model
\begin{equation}
\label{matrix1}
\frac{\partial G}{\partial \bar x}=\left(\begin{array}{cc}
I & I\\
B & (I + B)
\end{array}\right),
\end{equation}
where $\bar x$ is the $2N$-dimensional vector $\bar x \equiv (\bar p, \bar \theta)$, and
\begin{equation}
B=\left(\begin{array}{cccc}K_{\theta_1} & c_{21} & ... & c_{N1}\\
c_{12} & K_{\theta_2} & ...&  c_{N2}\\
\vdots & \vdots & \vdots &\vdots\\c_{1N} & c_{2N} & ... &
K_{\theta_N}\end{array}\right),
\end{equation}
with
\begin{equation}
K_{\theta_i}= a\cos[2\pi\theta_i(t)]+\frac{b}{\tilde N}\sum_
{j\neq
i}\frac{cos[2\pi(\theta_i(t)-\theta_j(t))]}{r^\alpha_{ij}},
\nonumber
\end{equation}
and
\begin{equation}
c_{ij}=c_{ji}= -\frac{b}{\tilde
N}\frac{cos[2\pi(\theta_i(t)-\theta_j(t))]}{r^\alpha_{ij}},
\nonumber
\end{equation}
where $i,j=1,\ldots , N$. It can be seen that,
\begin{equation}
\left(\frac{\partial G}{\partial \bar x}\right)^T=\left(\begin{array}{cc}
I & B\\
I & (I + B)
\end{array}\right),
\end{equation}
hence
\begin{equation}
 \left(\frac{\partial G}{\partial \bar x}\right)^T J = \left(\begin{array}{cc}
-B & I\\
 -(I + B) & I
\end{array}\right) \,.
\end{equation}
This quantity, multiplied (from the right side) by the matrix (\ref{matrix1})
yields $J$. Therefore our system is symplectic. Consequently, the $2N$
Lyapunov exponents $\lambda_1\equiv \lambda_M, \lambda_2,
\lambda_3,...,\lambda_{2N}$ are coupled two by two as
follows:$\lambda_1=-\lambda_{2N} \ge \lambda_2=-\lambda_{2N-1} \ge...\ge
\lambda_N=-\lambda_{N+1} \ge 0$. In other words, as a function of time, an
infinitely small {\it length} typically diverges as $e^{\lambda_1 t}$, an
infinitely small {\it area} diverges as   $e^{(\lambda_1 +\lambda_2)t}$, an
infinitely small {\it volume} diverges as $e^{(\lambda_1
  +\lambda_2+\lambda_3)t}$, an infinitely small $N$-dimensional {\it
  hypervolume} diverges as $e^{(\sum_{i=1}^N\lambda_i )t}$
($\sum_{i=1}^N\lambda_i $ being in fact equal to the Kolmogorov-Sinai entropy
rate, in agreement with the Pesin identity),  an infinitely small
$(N+1)$-hypervolume diverges as $e^{(\sum_{i=1}^{N-1}\lambda_i )t}$, and so
on. For example, a $(2N-1)$-hypervolume diverges as $e^{\lambda_1 t}$, and
finally a $2N$-hypervolume remains constant, thus recovering the conservative
nature of the system (of course, this corresponds to the Liouville theorem in
classical Hamiltonian dynamics).

For $\alpha=0$, similar models exist in the literature  though in different
contexts~\cite{moyanoetal,ahlersetal,kanekokonishi}. The present particular choice for the
coupling was made with the purpose of comparison with many-body Hamiltonian systems. Indeed, one can
derive the map set of equations (1) by applying a discretisation procedure to the
$\alpha$-XY model with an external field (for more details see~\cite{kanekokonishi,ott}).
As a consequence of having $N-1$ terms in the coupling summation
and the fact that there are $N$ maps, the simulation times are of order $O(N^2)$.
For this reason it is a difficult task to numerically simulate (\ref{eq_azp}) for
large values of $N$. To overcome this problem, we used an algorithm that takes
advantage of the symmetry of the lattice~\cite{firporuffo} and shortens the simulation
time to $O(N \ln N)$.

Initial coordinates and momenta were randomly taken from the following uniform distributions:
$\theta_i \in [\theta_0-\delta \theta, \: \theta_0+\delta \theta]$ and $p_i\in
[p_0-\delta p, \: p_0+\delta p]$.
For the coordinates we used $\theta_0=\delta \theta=0.5$ (i.e., $\theta_i \in
[0,1]$, {\em homogeneous} coordinate initial conditions).
For the momenta, we concentrated in two cases. To study the sensibility to initial
conditions we used a uniform distribution over the whole phase-space
($p_0=0.5$ and $\delta p=0.5$).
In the analysis of the relaxation to equilibrium we used a thin {\em waterbag}
intial condition with $p_0=0.3$ and $\delta p=0.05$.
We also checked {\em inhomogeneous} initial conditions, in particular
$p_0=\theta_0=0.3$ and $\delta p = \delta \theta=0.05$ (not shown in this paper). Note
that in this case there is no traslational symmetry in the
coordinates. For sufficiently long times, our simulations yield the
same sensitivity to initial conditions that we obtain for the symmetric case $\theta_i \in [0,1]$.
On the other hand, this inhomogeneous initial condition in the coordinates has an
important influence in the shape of the relaxation to equilibrium. This will
be further commented in the next section. We
note that the systematic study of the role of initial conditions is very
instructive,  but it is out of the scope of the present work.

\section{Results}
\label{results}

\subsection{Sensitivity to initial conditions}

In order to analyse the sensibility to initial conditions, we
numerically studied the largest Lyapunov exponent (LLE) for
different values of parameters $a$, $b$, $\alpha$ and $N$. We used the well-known method
developed by Benettin et al~\cite{benettinetal}.
As a consequence of the symplectic structure of (\ref{eq_azp}), the Lyapunov spectrum in the
$2N$-dimensional phase space of the map is, as already discussed, characterised by $N$ pairs of
Lyapunov coefficients, where each element of the pair is the negative of the
other. Therefore, the LLE sets an upper bound for the absolute value of the
entire spectrum of exponents.

We concentrated on the evolution of the LLE for different
values of $N$ starting with $\theta_0=0.5$, $\delta\theta=0.5, p_0=0.5$ and
$\delta p=0.5, \: \forall \, \alpha$. In a typical (sufficiently long)
realisation, the finite-time LLE, $\lambda_M$,
is a good estimator for the analytical
definition of the LLE (both quantities will coincide when $t\rightarrow
\infty$~\cite{anteneodo}). We averaged between realisations
(typically 100) in order to have small statistical fluctuations. We checked
that, for appropriately long times, $\lambda_M$ does not depend on the initial conditions
\cite{oseledec}.

\begin{figure}[ht!]
\begin{center}
  \includegraphics[width=0.75\columnwidth,angle=-90]{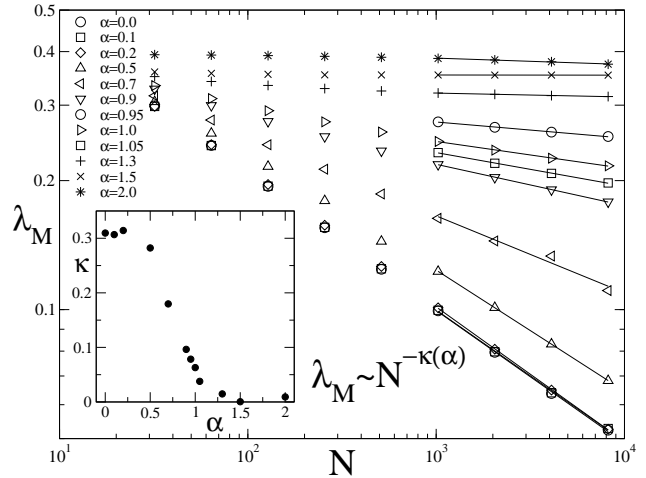}
\end{center}
\caption{Lyapunov exponent dependence on system size $N$ in log-log plot,
 showing that $\lambda_M\sim N^{-\kappa(\alpha)}$. Initial conditions
 correspond to $\theta_0=0.5$, $\delta \theta=0.5$, $p_0=0.5$ and $\delta
 p=0.5$. Fixed parameters are $a=0.005$ and
 $b=2$. We averaged over 100 realisations. {\it Inset:}  $\kappa$ vs. $\alpha$,
 exhibiting weak chaos  in the limit $N \to \infty$ when $0 \le \alpha\lessapprox 1$.
}
\label{fig1}
\end{figure}

In Fig.~\ref{fig1} we show the $N$-dependence of $\lambda_M$ for typical values of the interaction-range parameter
$\alpha$, and fixed values of parameters $a=0.005,\,
b=2$. Our results show that the value of $\lambda_M$ vanishes with increasing value of
$N$ (consequently, so does the rest of the Lyapunov spectrum)
as a power-law $\lambda_M\sim N^{-\kappa(\alpha)}$ for $\alpha\lessapprox 1$, and is
a positive constant for $\alpha>1$ ($\kappa\approx 0$). In the inset we
detail $\kappa$ as a function of $\alpha$. This shows that the map system
is {\em weakly chaotic} for long-range coupling
($\lambda_M\rightarrow0$ when $N\rightarrow\infty$), whereas
for short-range interactions, $\lambda_M$ remains positive for all $N$, meaning {\it strongly chaotic} dynamics (as
expected~\cite{zaslavsky,chirikov}).
Interestingly, this result is totally similar to the one numerically measured and analytically
predicted for the $\alpha$-XY model~\cite{firpo}, thus suggesting equivalent
behaviours. Indeed, and as stated by Anteneodo and Vallejos, this scaling is typical of
systems with couplings of the form
$1/r^{\alpha}$~\cite{anteneodovallejos}. Preliminary simulations suggest
that the fact that the weak chaos region extends slightly over $\alpha=1$ is an expected
consequence of finite-size and finite-time effects.

\begin{figure}[ht!]
\begin{center}
  \includegraphics[width=0.75\columnwidth,angle=-90]{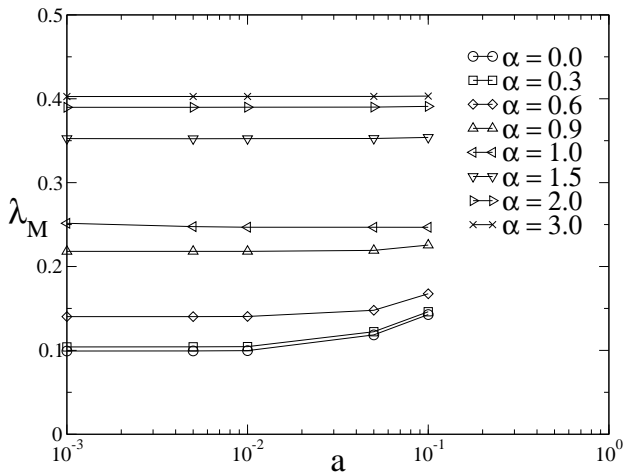}
\end{center}
\caption{Lyapunov exponent dependence on $a$ for different values of $\alpha$. Fixed constants are
  $N=1024$ and $b=2$. Initial conditions correspond to $\theta_0=0.5$,
  $\delta \theta=0.5$, $p_0=0.5$ and $\delta p=0.5$. We averaged over 100 realisations.
}
\label{fig2}
\end{figure}

The dependence of $\lambda_M$ with the nonlinear
parameter $a$ for different ranges of the interaction $\alpha$ is shown in
Fig. \ref{fig2}. We can see that, for $\alpha<1$, $\lambda_M$ decreases with $a$ and saturates for
$a\ll1$. This illustrates the influence of this nonlinear term. For increasing $a$, the sensibility to initial conditions
raises. On the other hand $a$ has almost no effect when $\alpha>1$,
where the $\lambda_M$ is approximately constant on the whole $a$-range. We
verified that for $a>1$ a slight increase appears for $\alpha>1$ (as in the long-range
case), but this effect is negligible.

\begin{figure}[ht!]
\begin{center}
\includegraphics[width=0.75\columnwidth,angle=-90]{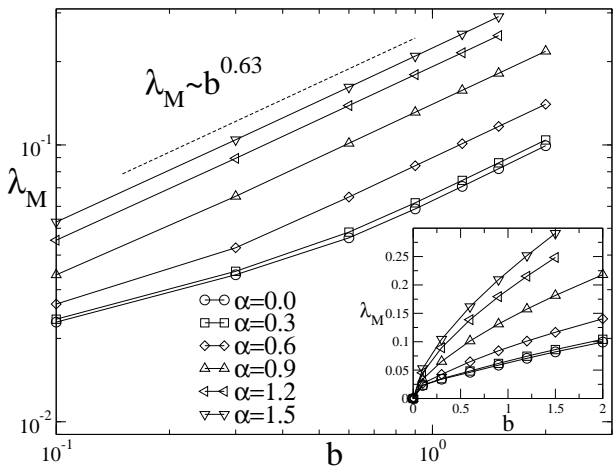}
\end{center}
\caption{Lyapunov exponent dependence on $b$ in log-log plot. Fixed constants are
  $N=1024$ and $a=0.005$. Initial conditions correspond to $\theta_0=0.5$,
  $\delta \theta=0.5$, $p_0=0.5$ and $\delta p=0.5$. We averaged over 100
  realisations. {\it Inset:} Same data in linear-linear plot.
}
\label{fig3}
\end{figure}

In Fig. \ref{fig3} we exhibit the dependence on the coupling parameter $b$,
also varying $\alpha$. For high values of $b$ ($b\gg1$), $\lambda_M = c(\alpha)\,
b^{\,0.63}\;\forall\alpha$, being $c(\alpha)$ a nonlinear function. Preliminary
simulations show that this exponent
varies only slightly with $a$ and $N$, suggesting that it may be unique. In
the case $b=0$, we verified that $\lambda_M = 0, \,
\:\forall \alpha$. For $\alpha>1$ this scaling is valid even for
$b\ll1$. On the other hand, if $\alpha<1$, a deviation from this power-law
behaviour emerges when $b\ll1$.

Our characterisation illustrates the fact that, quite generically, the
sensibility to initial conditions is strongly governed by the range of the
interactions. Orbits are clearly chaotic in the short-range case (as expected)
and strong numerical evidence of vanishing LLE emerges when interactions are long-range,
revealing a weakly chaotic regime.

Our present results for the maximal Lyapunov exponent can be approximatively
summarised as follows:

\begin{equation}
\lambda_M(a,b,N,\alpha) \propto f_{\alpha}(b)\,N^{-\kappa(\alpha)},
\end{equation}
where $0 \le a <<1$; $0 \le b \lessapprox 2$; $N>>1$, $\alpha \ge 0$, and $f_{\alpha}(b)$ is some function of $(b,\alpha)$ (e.g., $f_{\alpha}(b) \propto b^{0.63}$ for $\alpha>1$).

\subsection{Temperature evolution}

As stated above, system (\ref{eq_azp}) is symplectic, hence (hyper)volume
preserving, like Hamiltonian systems. A consequence of this is that
$\theta$ may be interpreted as a
``coordinate'' variable and $p$ as the conjugate ``momentum''.
We may define a  concept analogous to a {\it temperature} as twice the
mean ``kinetic energy" per particle~\cite{baldovinbrigatti,moyanoetal},

\begin{equation}
T(t)\equiv
\frac{1}{N}\sum_{i=1}^{N} \left(\langle
p_i^2(t)\rangle-\langle p_i(t)\rangle^2\right),
\label{temperature_standard}
\end{equation}
where $\langle ... \rangle$ denotes an ensemble average. This quantity can be interpreted as a dynamical analog, and plays a role similar to the
 physical temperature.
We refer to as the  {\it BG-temperature}, $T_{BG}$, the temperature associated with an
{\it uniform} ensemble distribution in phase space. This quantity may be analytically calculated,

\begin{equation}
\label{TBG}
T_{BG}\equiv \frac{1}{N}\sum_{i=1}^{N}\left[ \int_0^1
dp_i\;p_i^2-\left(\int_0^1dp_i\;p_i\right)^2 \right],
\end{equation}
which yields $T_{BG}=1/12\simeq 0.083\;(\forall N)$.

We studied the evolution of the dynamical temperature $T$ for typical values
of the parameters as described above, focusing on
the relaxation towards $T_{BG}$. We used {\em waterbag} initial
conditions with {\it homogeneous} coordinate distribution $\theta_i \in [0,1]$, and momenta centered at $p_0=0.3$
with width $\delta p=0.05$. The manner in which the relaxation takes
place depends strongly on the initial conditions. For example, using
$0<p_0\ll1$ makes the system to temporarily reach a temperature greater than
$T_{BG}$, as a consequence of the periodic boundary conditions (torus (${\rm
  mod}\;1$)). In general different initial conditions produce different relaxation
shapes, but scaling with $N$ remains similar. We also checked {\it inhomogeneous}
coordinate distributions as initial
conditions ($\theta_0=0.3,\: \delta \theta=0.05$), and obtained results qualitatively similar to those corresponding to
the homogeneous case we present here. Although the temperature value in the
metastable state is not as low as in the homogeneous case,
long-lasting metastable plateaux appear (very much as in the
$\alpha=0$ case of the $\alpha$-XY model, also known as Hamiltonian mean
field (HMF) for initial magnetisation $M=1$~\cite{pluchino}). The duration of these plateaux also scales
with $N$ in a qualitatively similar manner as for the homogeneous case. The
particular choice of initial conditions used in this work yields a rather
smooth relaxation with, for example, only one inflexion point, thus simplifying our analysis. The whole scenario is consistent with the conjecture advanced in ~\cite{conj}.

This type of relaxation has already been reported
for the particular case $\alpha=0$~\cite{moyanoetal}. A two-step process
appears: firstly a stage where $T<T_{BG}$, and then a final relaxation to
the predicted temperature $T_{BG}$. The initial regime varies very slowly
in time yielding quasistationary (QS) states.

\begin{figure}[!ht]
\begin{center}
  \includegraphics[width=0.75\columnwidth,angle=-90]{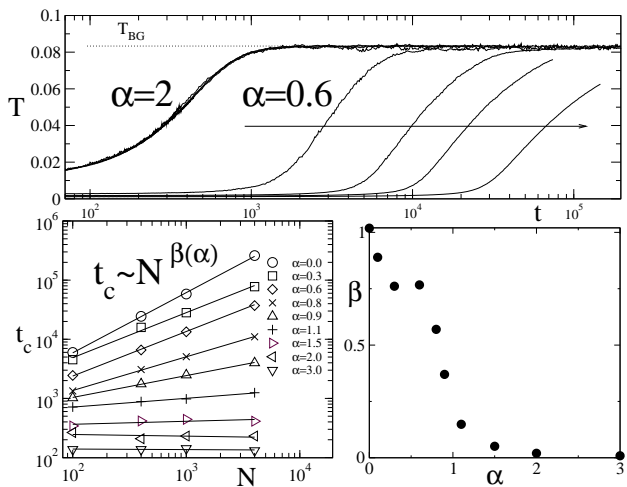}
\end{center}
\caption{{\it Upper panel:} Temperature evolution for $\alpha=2$ and $\alpha=0.6$ and four
  system sizes $N=100, 400, 1000, 4000$. Initial conditions correspond to $\theta_0=0.5$,
  $\delta \theta=0.5$, $p_0=0.3$ and
  $\delta p=0.05$. Fixed constants are $a=0.05$
  and $b=2$. For $\alpha=2$ the four curves coincide almost completely, all having
  a very fast relaxation to $T_{BG}$. For $\alpha=0.6$ the same sizes are
  shown, growing in the direction of the arrow. {\it Left bottom panel:} crossover
  time $t_c$ vs. $N$, showing a power-law dependence $t_c\sim\,N^{\beta(\alpha)}$ with
  $\beta(\alpha) \geq 0$. {\it Right bottom panel:} $\beta$ vs $\alpha$ shows that for long-range
  interactions the QS state life-time diverges in the thermodynamic
  limit. Note that when $\alpha=0, \,\beta=1$, and hence $t_c\propto N$. Given
  the non-neglectable error bars due to finite size effects, the relation
  $\beta=1-\alpha$ is not excluded as possibly being the exact one for $0 \le \alpha<1$ ($\beta=0$ otherwise); more
  precisely, it is not unplausible that $t_c \propto
  \frac{N^{1-\alpha}-1}{1-\alpha}$ (hence, for $\alpha=1$, $t_c \propto \ln N$).
}
\label{fig4}
\end{figure}

In Fig.~\ref{fig4} we show the temperature evolution for $\alpha=0.6$ and $\alpha=2$
for sizes $N=100,400,1000,4000$. The curves crossed by the arrow
correspond to the $\alpha=0.6$ case, being the first one (from left to right)
$N=100$. All four curves for $\alpha=2$ relax approximately at the same time,
so it has the appearance of a single curve. For $\alpha=0.6$ the typical two-step
relaxation is obtained.
From now on, we define the crossover time $t_c$ from the QS state to the BG equilibrium state
by means of the inflexion point, i.e. the time that corresponds to a maximum in the
time derivative of $T$. The dependence of $t_c$ with $N$
for this choice of parameters and initial conditions is reported in the
bottom left panel. The crossover time scales as $t_c\sim N^{\beta(\alpha)} \; \forall
\,\alpha$. For $\alpha \gtrapprox 1, \, \beta(\alpha)\approx 0$ and then $t_c$ remains constant
(as depicted for $\alpha=2$ in Fig. \ref{fig4}). For $\alpha \lessapprox 1,
\:\beta(\alpha) > 0$,  i.e., $t_c$ diverges in the thermodynamic limit $N\rightarrow\infty$. This
result indicates the inequivalence, for long-range interactions, of the orderings
$t\to\infty$ and then $N\to\infty$,  and $N\to\infty$ and then $t\to\infty$. Consistently, these QS
states become permanent (and therefore definitively relevant) when $N\to\infty$. Once again, the
situation that is found coincides with that of the $\alpha$-XY model~\cite{anteneodotsallis}.

\begin{figure}[ht!]
\begin{center}
  \includegraphics[width=0.95\columnwidth,angle=0]{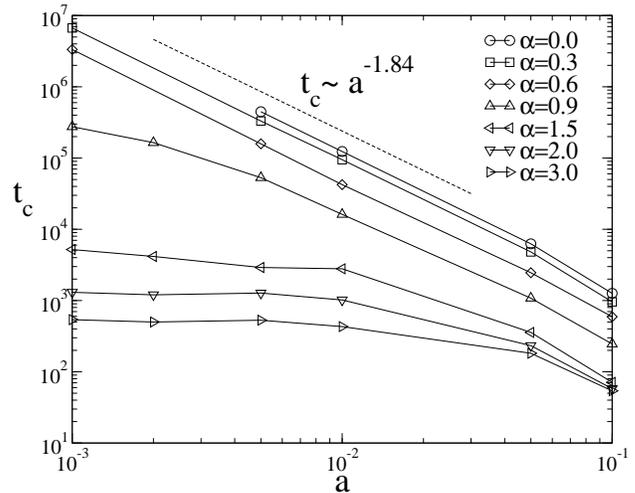}
\end{center}
\caption{Temperature dependence on $a$. Fixed constants are
  $N=100$ and $b=2$. Initial conditions correspond to $\theta_0=0.5$,
  $\delta \theta=0.5$, $p_0=0.3$ and $\delta p=0.05$. We averaged over 100 realisations.
}
\label{fig5}
\end{figure}

Finally, we studied the dependence of $t_c$ with the nonlinear parameter $a$,
for different values of $\alpha$, fixed coupling parameter $b=2$, and size $N=100$. In this
case, as well as in the other situations analised in this work, different
behaviours are reported for short- and long- range interactions. For
$\alpha>1$,  the value of $t_c$, in the limit $a\rightarrow 0$, tends to a
finite value. The situation is different in the long-range case,
where, for $\alpha\ll1$, the crossover time $t_c$ scales as
$t_c=d(\alpha)\:a^{1.84}$ being $d(\alpha)$ a nonlinear function. Preliminary
calculations show that the deviation for the case $\alpha=0.9$ and $a\ll1$ is
due to finite-size effects. This scaling law implies that, for
long-range interactions and vanishing nonlinear parameter $a$, the system stays
in the metastable regime permanently.

Our present results for the crossover time $t_c$ can be approximatively
summarised as follows:

\begin{equation}
t_c(a,b,N,\alpha) \propto
a^{-1.84}g_{\alpha}(b)\, N^{\beta(\alpha)},
\end{equation}
where $0.001 \le a <<0.1$, $0 \le b \lessapprox 2$, $N>>1$, $0 \le \alpha <1$, and
$g_{\alpha}(b)$ is some function of $(b,\alpha)$.

\section{Conclusions}
\label{conclusions}

In this work we have analysed a system of $N$ symplectically and globally coupled
standard maps, for both short- and long-range interactions. Our results connect
with studies made in long-range Hamiltonian
systems as well as with other map systems with vanishing largest Lyapunov exponent.
We studied a suitably defined dynamical temperature for a region of the
parameter space, and report the appearance of
long-lasting quasistationary states, followed by a relaxation to the predicted
equilibrium value. The relaxation time scales as $t_c\sim N^{\beta(\alpha)}$, with
$\beta(\alpha)\approx\,0$ when $\alpha\gtrapprox1$, and $\beta(\alpha)>0$ when
$0 \le \alpha\lessapprox\,1$, thus diverging as $N\rightarrow\infty$. This is a feature also present in the $\alpha$-XY Hamiltonian model, and constitutes a major conjecture in
nonextensive statistical mechanics (NSM).
Regarding the sensitivity to initial conditions, we calculated the maximum
Lyapunov exponent $\lambda_M$ as a function of the different system
parameters, namely, the size of the system $N$, the nonlinear
parameter $a$ and the coupling constant $b$.
We found that, as the number $N$ of maps grows, $\lambda_M$ vanishes as
$N^{-\kappa(\alpha)}$ with $\kappa(\alpha)>0$ for $0 \le \alpha\lessapprox\,1$, and
$\kappa\approx\,0$ for $\alpha\gtrapprox1$. The dependence of
$\kappa$ with $\alpha$ is reported and compared to be the same as in the
$\alpha$-XY model. We note that, even though the initial evolution of $\lambda_M$ depends strongly in the
initial conditions (as the temperature does), this scaling does not. In fact, it coincides with that
calculated with uniform initial conditions (i.e., initial $p_i,\theta_i$ uniformly
distributed in the $[0,1]$ interval) for $t\gg1$.
Our results exhibit that,  in the presence of long-range
interactions, the system tends to be {\em weakly chaotic} in the thermodynamic
limit. This is another feature also present in the nonextensive theory that
suggests its applicability.
The lack of ergodicity exhibited in our model is thought to be related to a
(multi)fractal constraint in the available phase-space that is enhanced in
certain limits such as $N\rightarrow\infty$. This possibility would be
connected to extremely low Arnold diffusion caused by the remanent of KAM tori
and islands, as has been suggested for similar high-dimensional systems~\cite{falcionietal}.
The similarities of the present coupled map model and the $\alpha$-XY Hamiltonian model suggest that both
share features that may have a common dynamical behaviour. This fact, together
with recent results obtained for low-dimensional maps, may clarify the role that NSM
plays in the correct description of long-range dynamical systems.

\section{acknowledgments}
Two of us (APM, LGM) warmly thank Pablo Barberis Blostein for
enlightening comments and discussions. LGM thanks Yuzuru Sato for useful
remarks, and CT thanks Ezequiel G.D. Cohen and Murray Gell-Mann for deeply stimulating
discussions. Financial support from CNPq (Brazilian agency), SECYT-UNC
(Argentinian agency) (APM), and SI International and AFRL (USA agencies) is
also acknowledged.


\begin{thebibliography}{99}
\bibitem{qexp}
The q-exponential is defined as $exp_q(x)\equiv [1+(1-q)\,x]^{1/1-q}$ and its
inverse is the q-logarithm, ${\rm ln}_q(x)\equiv (x^{1-q}-1)/(1-q)$. The usual
exponential and logarithm are recovered when $q\rightarrow 1$.
\end{thebibliography}
\end{document}